\documentclass[preprint,amsfonts,amssymb,amsmath,superscriptaddress]{revtex4}

	\usepackage{bm}

	\usepackage{feynmp}
 	\usepackage[pdftex]{graphicx}
	\newcommand{\vect}[1]{\boldsymbol{#1}}
	\newcommand{\op}[1]{\hat{\boldsymbol{#1}}}
	\newcommand{\px}{\tilde{p}_{x}}
	\newcommand{\py}{\tilde{p}_{y}}

	\newcommand{\strain}{\mathcal{A}}

	\newcommand{\req}[1]{Eq.~(\ref{#1})}
	\newcommand{\reqs}[1]{Eqs.~(\ref{#1})}
	\newcommand{\rref}[1]{(\ref{#1})}

	\newcommand{\beq}{\begin{equation}}
	\newcommand{\eeq}{\end{equation}}
	\newcommand{\be}{\begin{equation}}
	\newcommand{\ee}{\end{equation}}
	\newcommand{\beqa}{\begin{eqnarray}}
	\newcommand{\eeqa}{\end{eqnarray}}
	\newcommand{\bea}{\begin{eqnarray}}
	\newcommand{\eea}{\end{eqnarray}}
	
\begin{document}

\title{Landau levels in deformed bilayer graphene at low magnetic fields}
\author{Marcin Mucha-Kruczy\'{n}ski}
\email{m.mucha-kruczynski@lancaster.ac.uk}
\affiliation{Department of Physics, Lancaster University, Lancaster, LA1~4YB, United Kingdom}
\author{Igor L. Aleiner}
\affiliation{Physics Department, Columbia University, New York, NY 10027, USA}
\author{Vladimir I. Fal'ko}
\affiliation{Department of Physics, Lancaster University, Lancaster, LA1~4YB, United Kingdom}

\begin{abstract}
We review the effect of uniaxial strain on the low-energy electronic dispersion and Landau level structure of bilayer graphene. Based on the tight-binding approach, we derive a strain-induced term in the low-energy Hamiltonian and show how strain affects the low-energy electronic band structure. Depending on the magnitude and direction of applied strain, we identify three regimes of qualitatively different electronic dispersions. We also show that in a weak magnetic field, sufficient strain results in the filling factor $\nu=\pm 4$ being the most stable in the quantum Hall effect measurement, instead of $\nu=\pm 8$ in unperturbed bilayer at a weak magnetic field. To mention, in one of the strain regimes, the activation gap at $\nu=\pm 4$ is, down to very low fields, weakly dependent on the strength of the magnetic field.
\end{abstract}

\maketitle

\section{Introduction}
Electrons in monolayer graphene are chiral quasiparticles with Berry phase $\pi$, Dirac-like linear dispersion and Landau quantization in magnetic field leading to a peculiar spectrum with a Landau level at "zero energy" (the Fermi point separating conduction and valence bands in a neutral structure) \cite{geim_science_2009, novoselov_nature_2005, zhang_nature_2005, abergel_advphys_2010}. Electrons in bilayer graphene also exhibit an exotic behaviour. They can be viewed as `massive chiral fermions', that is, quasiparticles with a parabolic spectrum but also Berry phase of $2\pi$ \cite{geim_science_2009, abergel_advphys_2010, mccann_prl_2006, novoselov_natphys_2006}, and their high-magnetic-field spectrum features an eight-fold degenerate Landau level at zero energy. It has been noticed that the topology of the constant-energy lines in the dispersion bands in bilayer graphene may change, from single-connected, almost circular at higher energies, $\epsilon\sim 100$meV, into several disconnected at lower energies $\epsilon\lesssim 1$meV: a transformation known in the physics of metals as the Lifshitz transition \cite{lifshitz_jetp_1960}, which occurs when the Fermi level passes the saddle point in the spectrum. Theoretically, details of this transformation in the spectrum and the energy $\epsilon^{*}$ at which topology of the energy band changes are determined by the interplay between the nearest-neighbour and skew interlayer hoppings of electrons. For energies $|\epsilon|\ll |\epsilon^{*}|$, the single-particle electron spectrum consists of four Dirac cones with a linear dispersion, three with a Berry phase $\pi$ and one with $-\pi$. In this article, we show that the topology of the low-energy single-particle dispersion of bilayer graphene critically depends on mechanical deformations of the crystal. Strain not only influences the energy $\epsilon^{*}$ at which the topology of the dispersion changes but more critically, determines the number of Dirac cones in the low-energy part of BLG spectrum, from four in the unperturbed bilayer, down to two (both with the Berry phase equal to $\pi$) in a strongly strained crystal. To mention, no such dramatic changes occur in strained monolayer graphene where deformations of the crystalline lattice induce merely a shift in the position of the Dirac point in the momentum plane \cite{manes_prb_2007, pereira_prb_2009}. We also track the strain-induced spectral changes in the bilayer down to the evolution of the Landau levels for electrons, which enables us to predict what features in the quantum Hall effect in bilayer graphene crystals would persist down to the lowest magnetic fields and lowest carrier densities.

\begin{figure}[tbp]
\centering
\includegraphics[width=1.0\columnwidth]{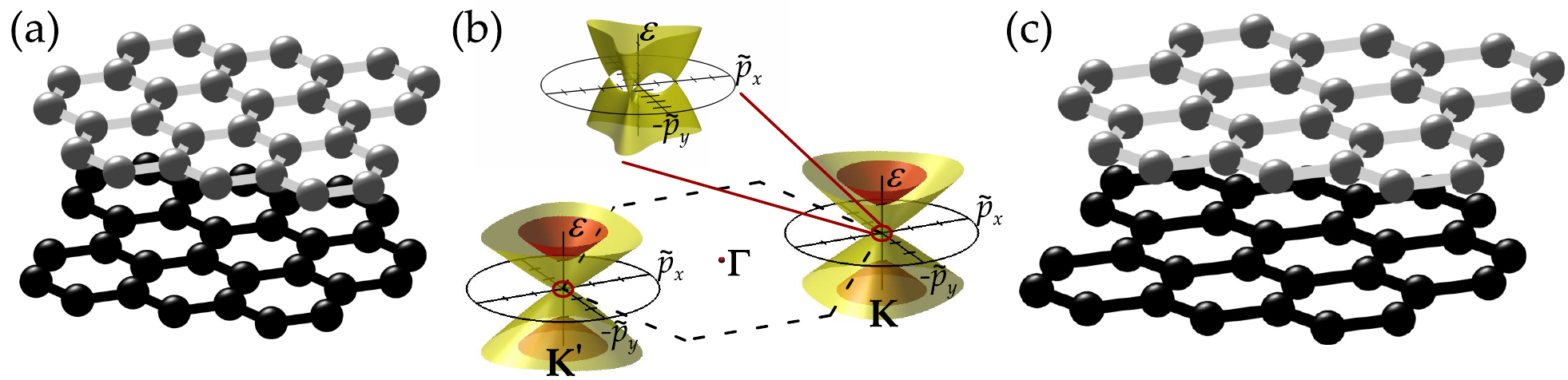}
\caption{(a) Side view of the unperturbed bilayer graphene lattice. The top (bottom) layer is shown in grey (black). (b) Electronic band structure of unperturbed bilayer graphene in the vicinity of the Brillouin zone corners $K$ and $K'$. In the top left is shown a detailed view of the low-energy band structure around the Lifshitz transition at energy $\epsilon^{*}$. (c) Side view of the deformed bilayer graphene lattice. The lattice was deformed using parameters $\delta=0.20$, $\delta '=-0.04$, $\theta=25^{\circ}$ and $\vect{\delta\!r}=\frac{r_{AB}}{4}(1,1)$.}
\label{fig:lattices}
\end{figure}

\section{Tight-binding model for strained bilayer graphene}
Bilayer graphene consists of two honeycomb layers of carbon atoms arranged according to Bernal stacking \cite{bernal_porsa_1924} (see Fig. \ref{fig:lattices}(a)) with four inequivalent carbon sites in the unit cell, $A_{1}$, $B_{1}$ in the bottom layer and $A_{2}$, $B_{2}$ in the top layer, and with $A_{2}$ positioned over $B_{1}$. Hexagonal symmetry in the plane leads to three nearest neighbours at the same distance surrounding each site, their position described with the help of vectors $\vect{e_{1}}=r_{AB}(0,1)$, $\vect{e_{2}}=r_{AB}(\frac{\sqrt{3}}{2},-\frac{1}{2})$ and $\vect{e_{3}}=r_{AB}(\frac{\sqrt{3}}{2},-\frac{1}{2})$, where $r_{AB}$ is the carbon-carbon distance. 

For the unperturbed bilayer, the electronic spectrum is gapless, with the conduction and valence bands touching in the corners of the hexagonal Brillouin zone, denoted as the K points at the energy conventionally used as the $\epsilon=0$ reference
level, which also coincides with the Fermi level in undoped bilayer graphene. Two additional bands are split from the neutrality point by energy $\epsilon\approx\pm\gamma_{1}$ (see Fig. \ref{fig:lattices}(b)). At very low energies, $\epsilon\sim 1$meV, the isoenergetic line undergoes a topological transition as it splits from a single-connected line into four disconnected parts, each resulting from a separate Dirac cone, as shown in the top of Fig. \ref{fig:lattices}(b) .

In the suspended graphene devices \cite{du_natnano_2008, bolotin_ssc_2008, feldman_natphys_2009, weitz_science_2010, martin_prl_2010} deformations can develop in the atomically thin carbon membranes due to the motion of contacts as the sample is cooled down. Understanding the influence of deformations of bilayer graphene in particular on its electronic properties is, now, becoming a pressing matter, in order to distinguish strain effects in suspended devices from the spectral changes expected in the recently predicted low-temperature phase transitions induced by the electron-electron interactions in undoped bilayer \cite{castro_prl_2008, zhang_prb_2010, vafek_prb_2010, lemonik_prb_2010, nandkishore_prl_2010, kotov_arxiv_2010} which are being searched for in suspended bilayer graphene devices \cite{feldman_natphys_2009, weitz_science_2010, martin_prl_2010}. Below, we characterise strain using angle $\theta$ between the principal axis of strain tensor and coordinates in Fig. \ref{fig:lattices}, and its eigenvalues, $\delta$ and $\delta'$. Recently investigated interlayer shear deformations \cite{son_arxiv_2010} can be included by allowing a shift of the top layer with respect to the bottom one, by $\vect{\delta\!r}=(\delta\!x,\delta\!y)$. As in monolayers, stretching of the lattice changes its symmetry and makes the intralayer hops $A1$-$B1$ ($A2$-$B2$) dependent on the direction of the hop, from a single value, $\gamma_{0}$ in the Slonczewski-Weiss-McClure parametrisation \cite{slonczewski_physrev_1958, dresselhaus_advphys_1981}, to three different values for the hops along bonds $\vect{e_{1}}$, $\vect{e_{2}}$ and $\vect{e_{3}}$, $\gamma_{0}^{(n)}=\gamma_{0}\left[1+\eta_{0}\left(\frac{\delta'-\delta}{2}\frac{\vect{e_{n}}}{r_{AB}}\!\cdot\!\vect{l}+\frac{\delta+\delta'}{2}\right)\right]$, where $n=1$, $2$ or $3$, and $\vect{l}=(\sin 2\theta,\cos 2\theta)$ takes into account unilateral deformation while the second term, $\frac{\delta+\delta'}{2}$, accounts for the `hydrostatic' rescaling of the lattice period and leads only to a change in the value of Dirac velocity, $v\rightarrow v \left[1+ \eta_{0}\frac{\delta+\delta'}{2}\right]$. Here, $\eta_{0}=\frac{r_{AB}}{\gamma_{0}}\frac{\partial\gamma_{0}}{\partial r_{AB}}$ quantifies the change of the intralayer $A$-$B$ hopping upon the change of the distance of carbon atoms on the plane. To describe the effect of strain in bilayers, we need to account for additional changes in the interlayer hoppings. The direct $A_{2}$-$B_{1}$ interlayer coupling, $\gamma_{1}$, may be changed by shear, $\gamma_{1}\rightarrow\gamma_{1}+O\!(\delta\!r^{2})$, which has no bearing on the topology of electron bands at low energies. More importantly, strain and shear make the value of the skew interlayer coupling $\gamma_{3}$ dependent on the direction of the hop, leading, again, to three distinct couplings $\gamma_{3}^{(n)}=\gamma_{3}\left\{1+\left[\frac{\vect{e_{n}}}{r_{AB}}\!\cdot\!\left(\frac{\delta'-\delta}{2}\vect{l}-\frac{\vect{\delta\!r}}{r_{AB}}\right)+\frac{\delta+\delta'}{2}\right]\eta_{3}\right\}$, where $\eta_{3}=\frac{r_{AB}}{\gamma_{3}}\frac{\partial\gamma_{3}}{\partial r_{AB}}$, for the three hops indicated in Fig. \ref{fig:lattices} with black dashed lines.

The variation of the hopping elements in the tight-binding model by strain and the lower symmetry of the bilayer lattice result in modification of the dispersion relation for electrons at low energies. The latter is most convenient to capture using an effective $2\times 2$ Hamiltonian, which describes in details the pair of low-energy bands near the Dirac point in bilayer graphene corresponding to electron states located predominantly on the sublattices $A_{1}$ and $B_{2}$ (those coupled by the skew hop $\gamma_{3}$, but not by the direct hop $\gamma_{1}$). The calculation leading to such effective Hamiltonian includes two steps. One consists in the Schrieffer-Wolff transformation \cite{schrieffer_physrev_1966} projecting the four-band model onto the effective two-band model \cite{mccann_prl_2006}. The other takes into account that cancellation between the contributions of the $\gamma_{0}$ and $\gamma_{3}$ hops in directions $\vect{e_{1}}$, $\vect{e_{2}}$ and $\vect{e_{3}}$ for the coupling of plane wave Bloch states on $A_{1/2}$ and $B_{1/2}$ sublattices in the corners $K$ and $K'$ of the Brillouin zone is no more exact. Note that lattice deformations also distort the hexagonal shape of the Brillouin zone, what is automatically taken into account in our analysis. As a result, the hopping integrals between sublattices are modified from $\xi v(p_{x}-i\xi p_{y})$ to $\xi v(p_{x}-i\xi p_{y})+\strain_{0}^{*}$ for the intralayer $A$-$B$ hops (where $v=\frac{a\sqrt{3}\gamma_{0}}{2\hbar}$ and $\strain_{0}=\frac{3}{4} (\delta-\delta') e^{-2i\theta} \gamma_{0}\eta_{0}$), and from $\xi v_{3}(p_{x}+i\xi p_{y})$ to $\xi v_{3}(p_{x}+i\xi p_{y})+\strain_{3}$ for the interlayer skew hops (where $v_{3}=\frac{a\sqrt{3}\gamma_{3}}{2\hbar}$ and $\strain_{3}=\frac{3}{4} (\delta-\delta') e^{-2i\theta} \gamma_{3}\eta_{3}-\frac{3}{2} \frac{\delta r}{r_{AB}} e^{i\varphi} \gamma_{3}\eta_{3}$ with $\varphi$ denoting the anticlockwise angle between negative direction of the $y$ axis and vector $\vect{\delta\!r}$). Then, the resulting two-band model can be described using the Hamiltonian
\begin{equation*}\label{eqn:Hamiltonian_unnumbered}
\op{H} =  -\frac{1}{\gamma_{1}}\left(\begin{array}{cc}
0 & \left(\xi v\op{\pi}^{\dagger} + \strain_{0}^{*}\right)^{2} \\
\left(\xi v\op{\pi} + \strain_{0}\right)^{2} & 0
\end{array}\right) + \left(\begin{array}{cc}
0 & \xi v_{3}\op{\pi} + \strain_{3} \\
\xi v_{3}\op{\pi}^{\dagger} + \strain_{3}^{*} & 0 
\end{array}\right),
\end{equation*}
where $m^{-1}=\frac{2v^{2}}{\gamma_{1}}$ and $\op{\pi}=p_{x}+ip_{y}$. Following the suggestion \cite{manes_prb_2007} that in monolayers the effect of homogeneous strain is equivalent to a constant vector potential, which can be eliminated by a gauge transformation equivalent to a shift of the valley centre from the Brillouin zone corners [$K$ ($\xi=+1$) or $K'$ ($\xi=-1$)], we introduce here valley momentum $\vect{\tilde{p}}=\vect{p}+\xi\frac{\hbar\eta_{0}}{r_{AB}}\frac{\delta-\delta'}{2}(\cos 2\theta,-\sin 2\theta)$, and arrive at the effective Hamiltonian for strained bilayer graphene:
\begin{align}\label{eqn:Hamiltonian}
\op{H} =  -\frac{1}{2m}\left(\begin{array}{cc}
0 & \left(\op{\tilde{\pi}}^{\dagger}\right)^{2} \\
\op{\tilde{\pi}}^{2} & 0
\end{array}\right) + \xi v_{3}\left(\begin{array}{cc}
0 & \op{\tilde{\pi}} \\
\op{\tilde{\pi}}^{\dagger} & 0 
\end{array}\right) + \left(\begin{array}{cc}
 0 & w \\
w^{*} & 0 
\end{array}\right),
\end{align}
where
\begin{align}
w = \frac{3}{4}(\eta_{3}-\eta_{0})\gamma_{3}e^{-i2\theta}(\delta-\delta ')-\frac{3}{2} \gamma_{3}\eta_{3}e^{i\varphi}\frac{\delta r}{r_{AB}}, \,\, \op{\tilde{\pi}}=\tilde{p}_{x}+i\tilde{p}_{y}. \nonumber
\end{align}

\section{The influence of the electron-electron interaction on the strain-induced term in the low-energy Hamiltonian}
\begin{fmffile}{diagrams}

In this section, we investigate the influence of electron-electron interactions on the low-energy Hamiltonian of (strained) bilayer graphene using the renormalisation group (RG) approach.
The derivation of the RG equations is based on the analysis of the leading
logarithmic divergences in the perturbation theory and their subsequent resummation. It follows the line of Ref.~\cite{lemonik_prb_2010},
though in this paper we will not consider the renormalizations of the marginal short range interaction giving rise to the 
instability towards nematic state \cite{lemonik_prb_2010}.
As usual, the bookkeeping of the perturbation theory is greatly simplified by using the diagrammatic technique. 

We pictorially represent the Hamiltonian \eqref{eqn:Hamiltonian} as $-\op{H} =  \bigcirc +  \bigtriangleup + \square;$
\be
\begin{split}
\ \bigcirc=\frac{1}{2m}\left(\begin{array}{cc}
0 & \left(\op{\tilde{\pi}}^{\dagger}\right)^{2} \\
\op{\tilde{\pi}}^{2} & 0
\end{array}\right);\ \bigtriangleup = - \xi v_{3}\left(\begin{array}{cc}
0 & \op{\tilde{\pi}} \\
\op{\tilde{\pi}}^{\dagger} & 0 
\end{array}\right) ; \square = - \left(\begin{array}{cc}
 0 & w \\
w^{*} & 0 
\end{array}\right).
\end{split}
\label{hamiltonian}
\ee

In the energy region where the RG flow occurs, two last terms in \req{hamiltonian} can be considered as the perturbation
so the Green function has the form 
\be
 -\quad \begin{fmfgraph*}(20,5) \fmfpen{thick} \fmfleft{l1}
  \fmfright{r1}
   \fmf{plain_arrow,label=$\epsilon,,\vect{p}$}{l1,r1}
 \end{fmfgraph*}
\ =\hat{G}(\epsilon,\vect{p})=\frac{1}{i\epsilon+\bigcirc}.
\label{Gf}
\ee
The Coulomb interaction denoted by
\be
\begin{fmfgraph*}(20,5)
\fmfpen{thick} \fmfleft{l1}
 \fmfright{r1}
  \fmf{wiggly,label=$\omega,,\vect{q}$}{l1,r1}
\end{fmfgraph*}
\ = -\frac{2\pi e^2}{q};
\label{bare-Coulomb}
\ee 
is divergent at $q\to 0$ so all the terms in $1/N$ approximation should be summed which gives
\be
\begin{fmfgraph*}(15,5)
\fmfpen{thick} \fmfleft{l1}
 \fmfright{r1}
  \fmf{dbl_wiggly}{l1,r1}
\end{fmfgraph*}
\quad =
\quad 
\begin{fmfgraph*}(15,5)
\fmfpen{thick} \fmfleft{l1}
 \fmfright{r1}
  \fmf{wiggly}{l1,r1}
\end{fmfgraph*}
\quad + \quad
\begin{fmfgraph*}(45,5)
\fmfpen{thick} \fmfleft{l1}
 \fmfright{r1}
\fmf{wiggly}{l1,l10}
\fmf{phantom,tension=0.7}{r10,l10}
  \fmf{dbl_wiggly}{r10,r1}
\fmffreeze
\fmf{plain_arrow,right=0.6,width=thick}{l10,r10}
\fmf{plain_arrow,right=0.6,width=thick}{r10,l10}
\fmfv{decor.shape=circle,decor.size=3thick,decor.filled=30}{l10}
\fmfv{decor.shape=circle,decor.size=3thick,decor.filled=30}{r10}
\end{fmfgraph*}.
\label{screened-Coulomb}
\ee
The explicit calculation of the polarization operator yields
\be
\begin{fmfgraph*}(9,5) \fmfpen{thin} 
\fmfleft{l1}
 \fmfright{r1}
\fmf{plain_arrow,right=0.6,width=thick}{l1,r1}
\fmf{plain_arrow,right=0.6,width=thick}{r1,l1}
\fmfv{decor.shape=circle,decor.size=3thick,decor.filled=30}{l1}
\fmfv{decor.shape=circle,decor.size=3thick,decor.filled=30}{r1}
\fmflabel{$\omega,q$}{r1}
\end{fmfgraph*}
\qquad
=\Pi(q,\omega)=\frac{Nm}{\pi D\left(\frac{2m \omega}{
      q^2}\right)};
\quad D(x)=\left[\ln\left(\frac{4x^2+4}{4x^2+1}\right)+\frac{2\arctan
    x -\arctan (2x)}{x}\right]^{-1},
\label{Pi}
\ee
where $N=4$ is the total degeneracy of the electronic spectrum. All the calculation is controlled in $1/N$ approximation. Using \reqs{screened-Coulomb} and \rref{Pi}, we obtain
\be
\begin{fmfgraph*}(15,5)
\fmfpen{thick} \fmfleft{l1}
 \fmfright{r1}
  \fmf{dbl_wiggly}{l1,r1}
\end{fmfgraph*}
\quad = -\frac{1}{\frac{q}{2\pi e^2} +\Pi }= - =\frac{\pi D\left(\frac{2m \omega}{
      q^2}\right)}{Nm};
\label{D}
\ee
The last equation is written under assumption that $qa_B\ll 1$ ( $a_B=1/Nme^2$ being the screening radius), which is always true in the region of applicability
of the Hamiltonian (1) of the main text. 

To derive the RG equations we, first, calculate the correction to the self-energy. 
On each step of the RG the integration over $\omega$ along the interaction line is restricted
by $E- \delta E < |\omega| < E$.
Its energy derivative gives the correction to the quasiparticle weight
\be
\begin{fmfgraph*}(25,10) \fmfpen{thin} 
\fmfleft{l10}
 \fmfright{r1}
\fmf{phantom,tension=12}{l10,l11}
\fmf{plain,width=thick,tension=7}{l11,l1}
\fmf{plain_arrow,width=thick}{l1,r11}
\fmf{plain,width=thick,tension=7}{r11,r1}
\fmffreeze
 \fmf{dbl_wiggly,left=0.6}{l1,r11}
\fmfv{decor.shape=circle,decor.size=4thick,decor.filled=30}{r11}
\fmfv{decor.shape=circle,decor.size=4thick,decor.filled=30}{l1}
\fmflabel{$
\displaystyle{\delta Z=\frac{i\partial}{\partial\epsilon}}$}{l10}
\fmflabel{$p=0$}{r1}
\end{fmfgraph*}
\label{Z}
\ee
which can be always eliminated by the rescaling of the Fermionic operators and thus does not affect by
itself any gauge invariant quantity. We will perform the rescaling in such a way that the quasiparticle weight is set to unity,
so that all the corrections have the meaning of the observable spectrum (note in passing that all the corrections to the observables
are logarithmic whereas $\delta Z$ is divergent more severely).
As a bonus, the scalar vertex is also not renormalized due to the gauge invariance
\[
\delta\quad
  \begin{fmfgraph*}(10,7) 
\fmfpen{thin} 
\fmftop{h}
\fmfbottom{b1,b2}
\fmf{plain_arrow,width=thick}{b1,b12}
\fmf{plain_arrow,width=thick}{b12,b2}
\fmfv{decor.shape=circle,decor.size=4thick,decor.filled=30}{b12}
\fmffreeze
\fmf{dbl_wiggly}{h,b12}
\end{fmfgraph*}
\quad
=
\delta Z\ \times\
 \begin{fmfgraph*}(10,7) 
\fmfpen{thin} 
\fmftop{h}
\fmfbottom{b1,b2}
\fmf{plain_arrow,width=thick}{b1,b12}
\fmf{plain_arrow,width=thick}{b12,b2}
\fmfv{decor.shape=circle,decor.size=4thick,decor.filled=30}{b12}
\fmffreeze
\fmf{dbl_wiggly}{h,b12}
\end{fmfgraph*}
\quad + \quad
 \begin{fmfgraph*}(15,7) 
\fmfpen{thin} 
\fmftop{h}
\fmfbottom{b1,b2}
\fmf{plain_arrow,width=thick}{b1,b12}
\fmf{plain_arrow,width=thick}{b12,b2}
\fmfv{decor.shape=circle,decor.size=4thick,decor.filled=30}{b12}
\fmfv{decor.shape=circle,decor.size=4thick,decor.filled=30}{b1}
\fmfv{decor.shape=circle,decor.size=4thick,decor.filled=30}{b2}
\fmffreeze
\fmf{dbl_wiggly}{h,b12}
\fmf{dbl_wiggly,right=0.6}{b1,b2}
\end{fmfgraph*}
\quad {+} \quad \raisebox{-0.6cm}{
\begin{fmfgraph*}(22,20) 
\fmfpen{thin} 
\fmftop{h}
\fmfbottom{b1,b2}
\fmf{plain_arrow,width=thick}{b1,b2}
\fmf{dbl_wiggly}{h,vh}
\fmf{dbl_wiggly,tension=0.7}{b1,v1}
\fmf{dbl_wiggly,tension=0.7}{b2,v2}
\fmf{plain_arrow,width=thick,tension=0.4}{vh,v2,v1,vh}
\fmfv{decor.shape=circle,decor.size=4thick,decor.filled=30}{vh}
\fmfv{decor.shape=circle,decor.size=4thick,decor.filled=30}{b1}
\fmfv{decor.shape=circle,decor.size=4thick,decor.filled=30}{b2}
\fmfv{decor.shape=circle,decor.size=4thick,decor.filled=30}{v1}
\fmfv{decor.shape=circle,decor.size=4thick,decor.filled=30}{v2}
\end{fmfgraph*}}
\quad {+} \quad\raisebox{-0.6cm}{
\begin{fmfgraph*}(22,20) 
\fmfpen{thin} 
\fmftop{h}
\fmfbottom{b1,b2}
\fmf{plain_arrow,width=thick}{b1,b2}
\fmf{dbl_wiggly}{h,vh}
\fmf{dbl_wiggly,tension=0.7}{b1,v1}
\fmf{dbl_wiggly,tension=0.7}{b2,v2}
\fmf{plain_arrow,width=thick,tension=0.4}{vh,v1,v2,vh}
\fmfv{decor.shape=circle,decor.size=4thick,decor.filled=30}{vh}
\fmfv{decor.shape=circle,decor.size=4thick,decor.filled=30}{b1}
\fmfv{decor.shape=circle,decor.size=4thick,decor.filled=30}{b2}
\fmfv{decor.shape=circle,decor.size=4thick,decor.filled=30}{v1}
\fmfv{decor.shape=circle,decor.size=4thick,decor.filled=30}{v2}
\end{fmfgraph*}}
\quad{=0}
\]

The correction to the effective mass is found from 
\be
\delta\bigcirc = \delta Z \times \bigcirc
+
\quad
\raisebox{-15pt}{
\begin{fmfgraph*}(25,10) \fmfpen{thin} 
\fmfleft{l4}
 \fmfright{r}
\fmflabel{$\epsilon=0$}{r}
\fmf{plain,width=thick,tension=7}{l4,l40}
\fmf{plain_arrow,width=thick}{l40,r10}
\fmf{plain,width=thick,tension=7}{r10,r}
\fmffreeze
\fmf{dbl_wiggly,left=0.6}{l40,r10}
\fmfv{decor.shape=circle,decor.size=4thick,decor.filled=30}{r10}
\fmfv{decor.shape=circle,decor.size=4thick,decor.filled=30}{l40}
\end{fmfgraph*}
}
\qquad\quad
=\frac{\bigcirc}{N}  
\left(\frac{dE}{ E}\right)\int_{0}^\infty \frac{d x}{\pi}  \frac{D(x)(1-3x^2)}{(1+x^2)^3}.
\label{mass}
\ee

The other corrections are
\be
\begin{split}
&\delta \bigtriangleup
=
\delta Z\times \bigtriangleup
+ 
\raisebox{-15pt}{
\begin{fmfgraph*}(25,10) \fmfpen{thin} 
\fmfleft{l4}
 \fmfright{r}
\fmflabel{$\epsilon=0$}{r}
\fmf{plain,width=thick,tension=7}{l4,l40}
\fmf{plain_arrow,width=thick}{l40,c,r10}
\fmf{plain,width=thick,tension=7}{r10,r}
\fmffreeze
\fmf{dbl_wiggly,left=0.6}{l40,r10}
\fmfv{decor.shape=circle,decor.size=4thick,decor.filled=30}{r10}
\fmfv{decor.shape=circle,decor.size=4thick,decor.filled=30}{l40}
\fmfv{decor.shape=triangle,decor.size=7thick,decor.filled=empty}{c}
\end{fmfgraph*}
}
\qquad\quad=\frac{ \bigtriangleup}{N}
\left(\frac{dE}{E}\right)  
\int_{0}^\infty \frac{d x}{\pi}  \frac{D(x)(1-3x^2)}{(1+x^2)^3};
\\
&\delta \square =  \delta Z\times \square +
\raisebox{-15pt}{
\begin{fmfgraph*}(25,10) \fmfpen{thin} 
\fmfleft{l4}
 \fmfright{r}
\fmflabel{$\epsilon=0$}{r}
\fmf{plain,width=thick,tension=7}{l4,l40}
\fmf{plain_arrow,width=thick}{l40,c,r10}
\fmf{plain,width=thick,tension=7}{r10,r}
\fmffreeze
\fmf{dbl_wiggly,left=0.6}{l40,r10}
\fmfv{decor.shape=circle,decor.size=4thick,decor.filled=30}{r10}
\fmfv{decor.shape=circle,decor.size=4thick,decor.filled=30}{l40}
\fmfv{decor.shape=square,decor.size=5thick,decor.filled=empty}{c}
\end{fmfgraph*}
}
\qquad \quad
= \frac{\square}{N}
\left(\frac{dE}{E}\right)
\int_{0}^\infty \frac{d x}{2\pi}  \frac{D(x)}{(1+x^2)^2}.
\end{split}
\label{other}
\ee

Calculating the integrals in \reqs{mass}-\rref{other}, substituting $N=4$, and $d\lambda = d E/E$,
 we obtain:
\be\label{eqn:rg_results}
\partial_{\lambda}w = 0.11w, \,\, \partial_{\lambda}m^{-1} = -0.02m^{-1}, \,\, \partial_{\lambda}v_{3}=-0.02v_{3},
\ee
\end{fmffile}
where $\lambda = \ln\frac{\gamma_{1}}{\epsilon}$ and $\epsilon$ is the running energy scale. 
The electron-electron interaction enhances the strain-induced term stronger than other parameters, 
and at energies $\epsilon\sim|w|$, where the influence of strain plays a dominant role in 
determining the electron spectrum, we substitute 
$|w| \to |w|\exp(0.11\ln\frac{\gamma_{1}}{|w|})\approx |w|^{0.89}\gamma_{1}^{0.11}$, 
in the Hamiltonian \eqref{eqn:Hamiltonian}.

\begin{figure*}
\centering
\includegraphics[width=1.0\textwidth]{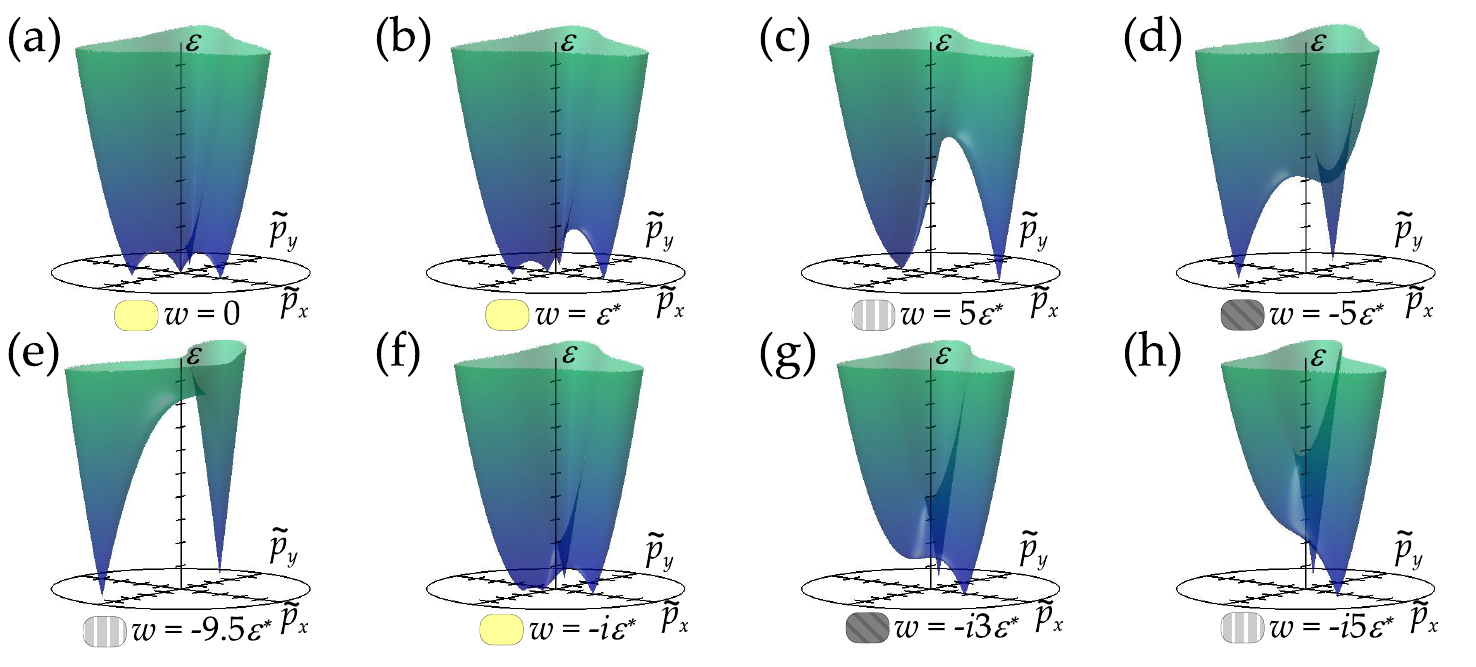}
\caption{Three-dimensional plots of the low-energy electronic dispersions in the conduction band of (strained) BLG, for representative points in the $(\Re w,\Im w)$ space. The energy and momentum are measured in the units of $\epsilon^{*}=mv_{3}^{2}/2$ and $mv_{3}$, respectively. The following values were used to obtain the graphs: the effective mass $m=0.035m_{e}$ ($m_{e}$ is the electron mass), $v_{3}=10^{5}$m/s. The resulting energy $\epsilon^{*}\approx 1$meV.}
\label{fig:dispersions}
\end{figure*}

\section{Low-energy electronic band structure of strained bilayer graphene}
\begin{figure}[tbp]
\centering
\includegraphics[width=0.5\columnwidth]{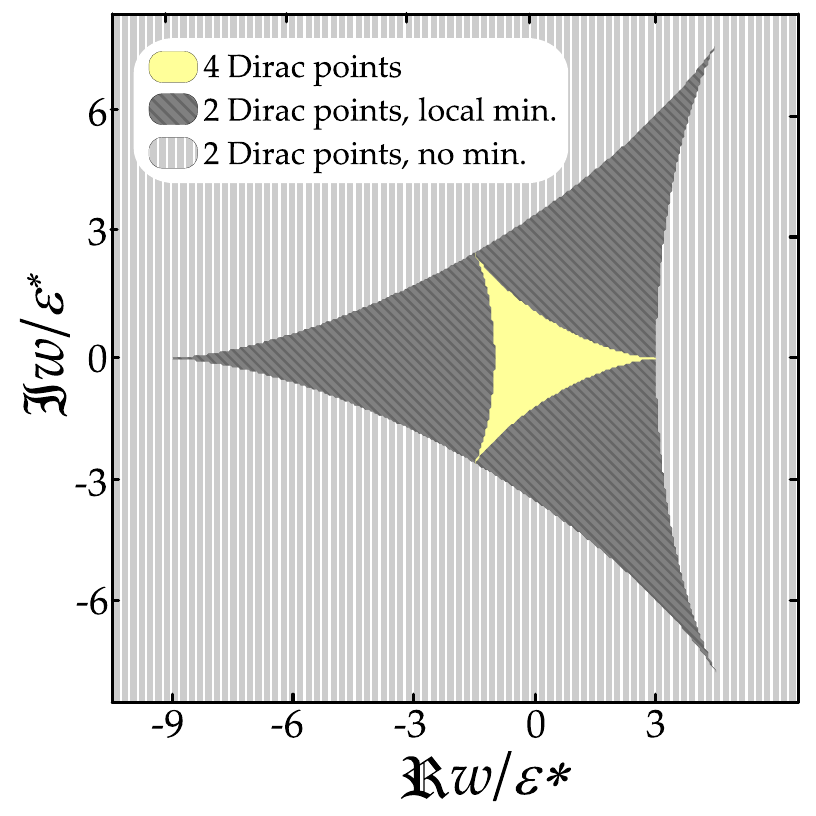}
\caption{Diagram showing what distinctive features are present in the low-energy electronic dispersion of BLG as a function of the deformation.}
\label{fig:phase_diagram}
\end{figure}

The resulting low-energy electronic dispersions of the conduction band [the band structure described by Hamiltonian in equation \eqref{eqn:Hamiltonian} is always electron-hole symmetric] of strained bilayer graphene in the vicinity of the valley $K_{+}$ are displayed for representative values of $(\Re w,\Im w)$ in Fig. \ref{fig:dispersions}. The dispersion for the case of the unperturbed bilayer, that is for $w=0$, is shown in Fig. \ref{fig:dispersions}(a). It is quasi-parabolic at energies $\epsilon\gg\epsilon^{*}=\frac{mv_{3}^{2}}{2}$, with $v_{3}$ responsible for the trigonal warping of isoenergetic lines. At $\epsilon=\epsilon^{*}$, the isoenergetic line undergoes a Lifshitz transition and splits into four Dirac cones \cite{mccann_prl_2006}. Remaining dispersions in the top row depict the bottom of the conduction band for $\Re w\neq 0$, $\Im w=0$. As the value of $\Re w$ is increased (Fig. \ref{fig:dispersions}(b)-(c)), the two side cones positioned off the $\px$ axis and the central cone move closer, as shown in the graph for $w=\epsilon^{*}$. Those three cones collide for $w=3\epsilon^{*}$, and for $w>3\epsilon^{*}$ only two Dirac cones remain, as shown in Fig. \ref{fig:dispersions}(c) for $w=5\epsilon^{*}$. If $\Re w<0$ is negative (graphs (d) and (e) in Fig. \ref{fig:dispersions}), the central cone and the side cone positioned on the $x$ axis approach each other and collide for $w=-\epsilon^{*}$, creating a local, quasi-parabolic minimum. This minimum persists for some range of the strain, although it lifts off the $\epsilon=0$ plane, as shown in the graph \ref{fig:dispersions}(d) for $w=-5\epsilon^{*}$. Eventually, for $\Re w=-9\epsilon^{*}$, the minimum merges with a saddle point and with further decrease of $\Re w$, again only two Dirac cones remain in the spectrum, as in Fig. \ref{fig:dispersions}(e) for $w=-9.5\epsilon^{*}$. A contrasting situation of $\Re w=0$, $\Im w\neq 0$, is presented in the graphs in Fig. \ref{fig:dispersions}(f)-(h) (due to symmetry, dispersions for opposite values of $\Im w$ are mirror reflections of each other with respect to the $\py=0$ plane and hence only situation of $\Im w<0$ is described here). Again, starting from the unperturbed system and increasing the magnitude of $\Im w$ leads to two of the cones moving closer to each other (Fig. \ref{fig:dispersions}(f) for $w=-i\epsilon^{*}$) and colliding with a creation of a local, quasi-parabolic minimum which lifts off the $\epsilon=0$ plane (graph (g) in Fig. \ref{fig:dispersions} for $w=-i3\epsilon^{*}$) and eventually disappears, leaving only two Dirac cones (graph (h) in Fig. \ref{fig:dispersions} for $w=-i5\epsilon^{*}$). We identify three qualitatively different regimes in the $(\Re w,\Im w)$ space: (i) spectrum contains four Dirac cones, (ii) spectrum contains two Dirac cones and a local minimum, created by a collision of two of the cones and which in general does not touch the $\epsilon=0$ plane, and (iii) spectrum contains only two Dirac cones. The extent of those regimes is shown in Fig. \ref{fig:phase_diagram}.

\begin{figure}[tbp]
\centering
\includegraphics[width=0.5\columnwidth]{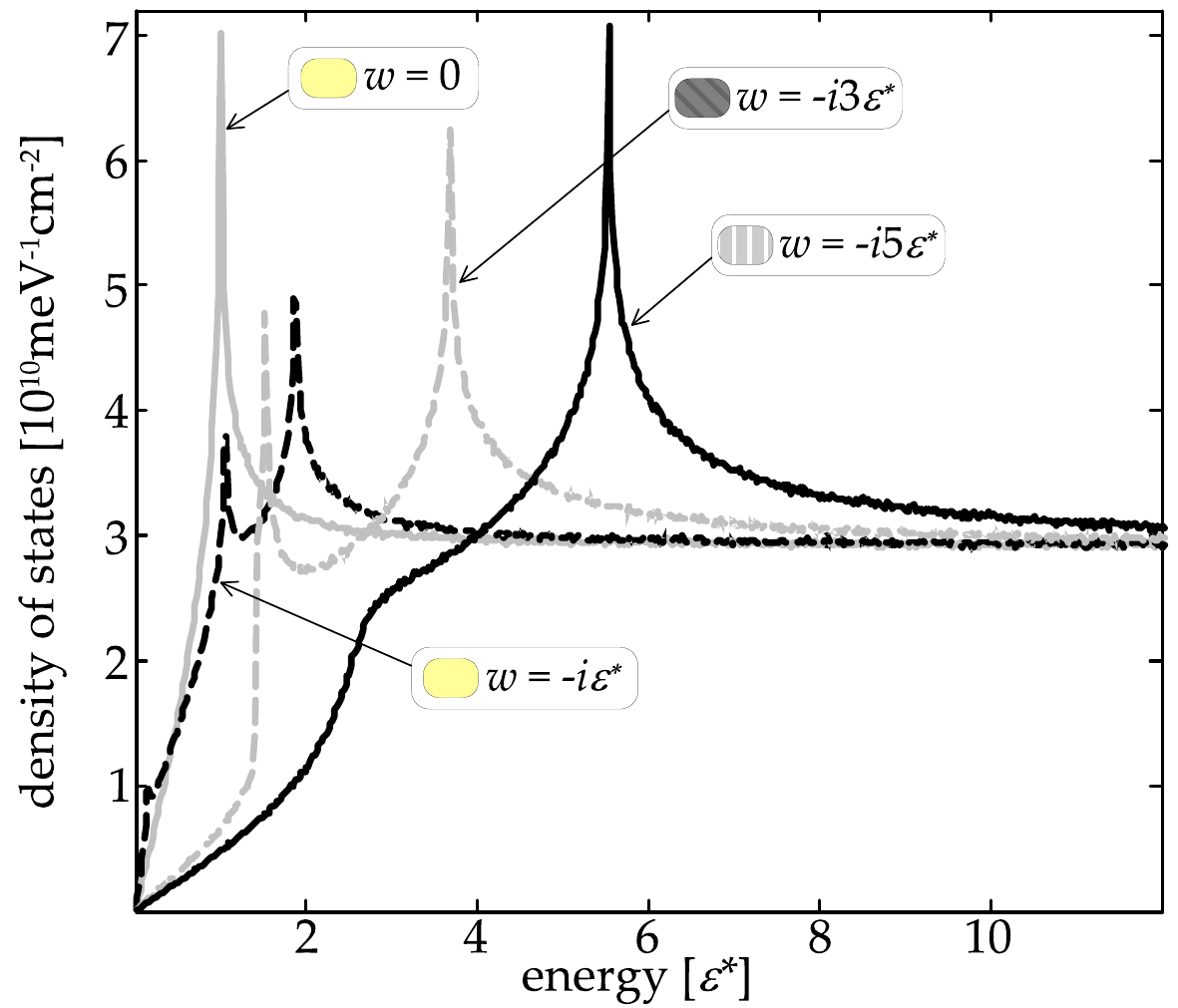}
\caption{Low-energy density of states in the conduction band of (strained) bilayer graphene for representative values of $w$.}
\label{fig:dos}
\end{figure}

Strain-induced deformation of the low-energy electronic dispersion gives rise to distinctive features in the low-energy density of states in each of the strain regimes. In particular, existence of a local parabolic minimum results in a sharp step in the density of states, while saddle points lead to van Hove singularities. Note that for $w\neq 0$, the symmetry between the side cones is broken and the saddle points are no longer all found at the same energy $\epsilon^{*}$, leading to more than one van Hove singularity and more than one Lifshitz transition. Densities of states representative of all the regimes are shown in Fig. \ref{fig:dos} for $w=0$,  $w=-i\epsilon^{*}$, $w=-i3\epsilon^{*}$ and $w=-i5\epsilon^{*}$, that is points that lie on the $\Re w=0$ line. Indeed, for $w=0$ only one van Hove singularity is present, whereas three can be traced for $w=-i\epsilon^{*}$. A sharp step, corresponding to the contribution of a quasi-parabolic part of the band structure to the density of states, is visible for $w=-i3\epsilon^{*}$. It is followed by a van Hove singularity, as for this dispersion the saddle point lies not far above the position of the local minimum. Finally, for $w=-i5\epsilon^{*}$, the sharp step is washed out (as the local minimum merged with the saddle point) and only one van Hove singularity remains.

\section{Landau levels of strained bilayer graphene}
\begin{figure}[tbp]
\centering
\includegraphics[width=1.0\columnwidth]{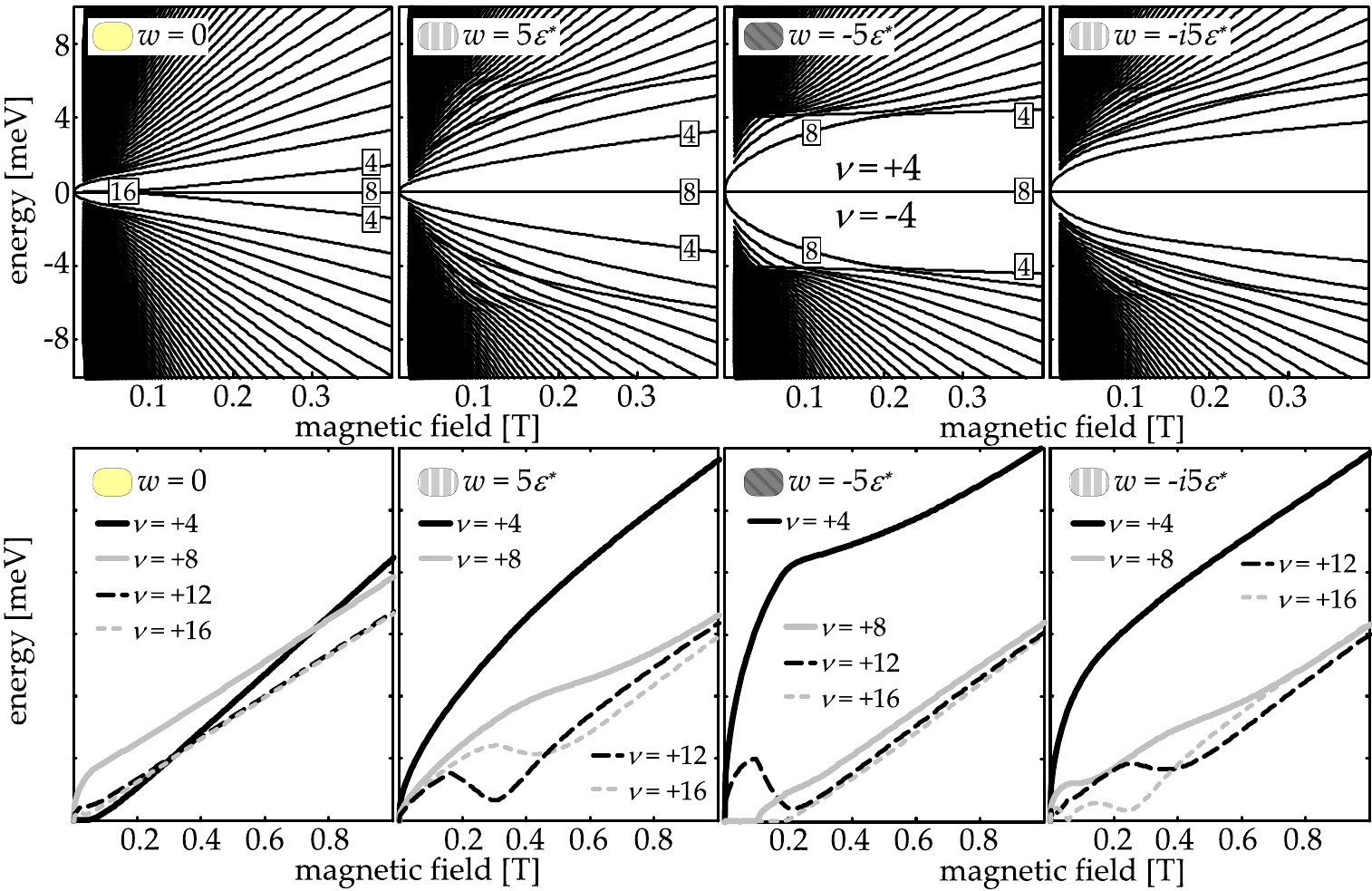}
\caption{Top row: fan plots of Landau levels for (strained) bilayer graphene, for representative points in the $(\Re w,\Im w)$ space. Numbers in boxes correspond to the additional degeneracy of a given Landau level (valley and spin included) and $\nu$ on some of the graphs denotes the most stable filling factor. Bottom row: activation energies for the quantum Hall effect in strained bilayer graphene with various integer filling factors. For a large enough strain, filling factor $\nu=\pm 4$ would be the only persistent feature in the low-field quantum Hall effect. The appearance of a local minimum in the electron dispersion upon a collision of two Dirac points ($w=-5\epsilon^{*}$) is manifested by an intermediate saturation of the magnitude of the activation gap within the interval $0.2T<B<0.4$T.}
\label{fig:LLs_and_act_gaps}
\end{figure}

Presence of strain in bilayer graphene may also lead to new features in the low-energy Landau level (LL) spectrum. To obtain Landau level spectra for bilayer graphene in a perpendicular external magnetic field, we use the two-band Hamiltonian \eqref{eqn:Hamiltonian}, where the electron momentum, $\vect{\tilde{p}}=-i\hbar\nabla-e\vect{A}$, incorporates the electromagnetic vector potential $\vect{A}$. For simplicity, we use the Landau gauge, $\vect{A}=(0,Bx)$. The operators $\op{\tilde{\pi}}$ and $\op{\tilde{\pi}}^{\dagger}$ fulfil the role of raising and lowering operators when acting on the magnetic oscillator functions $\psi_{n}\equiv e^{-iqy}\phi_{n}(x+q\lambda_{B}^{2})$, where the magnetic length $\lambda_{B}=\sqrt{\frac{\hbar}{eB}}$. If we neglect the second and third terms in the effective Hamiltonian (1), we can describe the low-energy Landau levels as a sequence of levels \cite{mccann_prl_2006, abergel_prb_2007, ando_jpsj_2007, koshino_prb_2008, nakamura_prb_2008, mucha-kruczynski_jpcm_2009}
\begin{equation}\label{eqn:two-band_model_LLs}
\!\!\!|n^{\alpha}\rangle\!=\!\!\left(\begin{array}{c} \!\psi_{n} \\ \!0 \end{array}\!\right)\!,~n\!=\!0,1,~~
|n^{\alpha}\rangle\!=\!\frac{1}{\sqrt{2}}\!\left(\begin{array}{c} \psi_{n} \\ \alpha\psi_{n-2} \end{array}\right)\!,~n\!\geq\!2,
\end{equation}
with energies $\epsilon_{n^{\alpha}}=\alpha\hbar\omega_{c}\sqrt{n(n-1)}$, where $\omega_{c}=eB/m$, and $\alpha$ distinguishes between the conduction ($\alpha=1$) and the valence ($\alpha=-1$) band. To incorporate the $v_{3}$ and the strain-induced terms into this picture, we follow the approach employed before to describe the effect of the $\gamma_{3}$ coupling on the LL structure of graphite \cite{nakao_jpsj_1976}. We describe the new eigenstates, each of them a linear combination of an infinite number of functions $\psi_{n}$, with vectors which $(2n-1)$th and $(2n)$th component create a minimal subspace required to describe the $n$th LL in the absence of $v_{3}$. In the corresponding infinite Hamiltonian matrix,
\begin{equation}
\op{H}= \left(\begin{array}{cccccc}
0 & 0 & \op{W} & \op{D}(1) & 0 & \cdots \\
0 & 0 & 0 & \op{W} & \op{D}(2) & \cdots \\
\op{W}^{\dagger} & 0 & \op{H}(1) & 0 & \op{W} & \cdots \\
\op{D}^{\dagger}(1) & \op{W}^{\dagger} & 0 & \op{H}(2) & 0 & \cdots \\
0 & \op{D}^{\dagger}(2) & \op{W}^{\dagger} & 0 & \op{H}(3) & \cdots \\
\vdots & \vdots & \vdots & \vdots & \vdots & \ddots 
\end{array}\right),
\end{equation}
where
\begin{align}
\op{H}(n) & = \frac{v^{2}}{\gamma_{1}^{2}}\hbar\omega_{c}\sqrt{n(n+1)} \left(\begin{array}{cc}
 0 & 1 \\
 1 & 0 
\end{array}\right), \\
\op{D}(n) & = \left(\begin{array}{cc}
 0 & -i\xi \frac{vv_{3}}{\gamma_{1}}\sqrt{2\hbar eBn} \\
 0 & 0 
\end{array}\right), \\
\op{W} & = \left(\begin{array}{cc}
 0 & w \\
 0 & 0 
\end{array}\right),
\end{align}
the first term in the low-energy Hamiltonian results in decoupled diagonal $2\times 2$ blocks, the second term results in some off-diagonal couplings between functions $\psi_{n}$ and $\psi_{n\pm 3}$, while the strain-induced third term leads to off-diagonal couplings between functions $\psi_{n}$ and $\psi_{n\pm 2}$ \cite{footnote}. We then truncate the infinite matrix, restricting the calculation to a given $n$ LLs and diagonalise the resulting Hamiltonian numerically.  The number of the basis vectors required in the calculation in order to describe correctly the low-energy LL structure increases with decreasing magnetic field (reflecting growing importance of the $v_{3}$ terms at weaker fields). In this work, matrices of the dimension 1200 have been used to produce the spectra, which are shown in the top row of Fig. \ref{fig:LLs_and_act_gaps} for $w=0$, $w=5\epsilon^{*}$, $w=-5\epsilon^{*}$ and $w=-i5\epsilon^{*}$.

The additional degeneracy of the $0$th LL in bilayer graphene at weak fields is connected to the number of Dirac cones in the electronic spectrum. At magnetic fields such that the inverse of the magnetic length $\lambda_{B}$ is smaller than the distance in the reciprocal space between a pair of cones, each of them provides additional four-fold degeneracy to the LL (just like in monolayer graphene \cite{novoselov_nature_2005, zhang_nature_2005}). At stronger fields, contributions of separate cones can no longer be resolved, leading to an 8-fold degenerate $0$th LL \cite{mccann_prl_2006, novoselov_natphys_2006}. Hence, for the strain $w$ within the region shown in yellow in Fig. \ref{fig:phase_diagram}, the zeroth LL becomes 16-fold degenerate at very weak but nonzero fields as three LLs, 8-, 4- and 4-fold degenerate, merge together, reflecting the existence of four Dirac cones in the spectrum. However, for regions of greater strain, shown in Fig \ref{fig:phase_diagram} with dark and light shading, the zeroth LL is 8-fold degenerate at all fields due to only two cones in the electronic spectrum. As the size of the gap between two LLs decreases with the LL index, the filling factor $\nu$ determining the biggest activation gap at low fields is closely connected to the additional degeneracy of the zeroth LL. Graphs showing the magnitude of activation gaps for filling factors $\nu=\pm 4$, $\nu=\pm 8$, $\nu=\pm 12$ and $\nu=\pm 16$ as a function of magnetic field for the values of $w$ representing all regions from Fig. \ref{fig:phase_diagram}, are shown in the bottom row in Fig. \ref{fig:LLs_and_act_gaps}.  For the region of weak strain, here portrayed with $w=0$, the biggest activation gap occurs at $\nu=\pm 8$, whereas for the two other regions ($w=5\epsilon^{*}$, $w=-5\epsilon^{*}$ and $w=-i5\epsilon^{*}$), filling factor $\nu=\pm 4$ is the most stable. For the region of intermediate strain ($w=-5\epsilon^{*}$), where a local parabolic minimum exists in the electronic dispersion, the activation energy between the $0$th and the first LL experiences a very unusual intermediate saturation, indicating that one of the LLs gets stuck in that minimum.

\section{Acknowledgements}

This project has been funded by the EPSRC grants EP/G041954 and Science \& Innovation Award EP/G035954. This review is based upon the lecture course "Electronic properties of graphene" taught by V. Fal'ko at the 5th Windsor Summer School, $9-21/8/2010$, in Windsor, United Kingdom \cite{windsor}.


\begin{thebibliography}{99}
\bibitem{footnote} Note, that two rows and columns in the above matrix, identically equal to zero, give rise to two solutions at zero energy, which correspond to unphysical eigenstates (zero vectors) and should not be confused with the zero energy Landau levels.
\end{thebibliography}
\end{document}